\documentstyle[times,pramana,epsf,floats]{ias}
\begin{document}
\mark{{Infra-red fixed points in supersymmetry}{B. Ananthanarayan}}
\title{Infra-red fixed points in supersymmetry}

\author{B. Ananthanarayan}
\address{Centre for Theoretical Studies, Indian Institute of Science, 
Bangalore 560 012, India}
\keywords{Supersymmetry, Renormalization Group, Fixed Points}
\pacs{11.10.Hi, 11.30.Fs, 12.60.Jv}
\abstract{Model independent constraints on supersymmetric models
emerge when certain couplings are drawn towards their infra-red 
(quasi) fixed points in the course of their renormalization group evolution.
The general principles are first reviewed and the conclusions for some
recent studies of theories with R-parity and baryon and lepton number
violations are summarized.}

\maketitle
\section{Introduction}
The Standard Model (SM) is a great success in describing
the strong and electroweak interactions based on an underlying gauge
principle, but one of its main weaknesses is that
the masses of the matter particles, the quarks and leptons,  are
free parameters of the theory.  This weakness persists in 
the Minimal Supersymmetric Standard Model (MSSM).  
The fermions mass problem in the MSSM arises from the 
presence of many unknown dimensionless Yukawa couplings.  
On the other hand the MSSM leads to a successful prediction
for the ratio of the gauge couplings with a gauge unificiation scale 
$M_G \simeq 10^{16}$ GeV.  It, therefore, becomes important
to perform the radiative corrections in determining all the dimension
$\leq 4$ terms in the lagrangian.  This can be achieved by using the 
renormalization group equations in finding the values of parameters
at the low scale, given their value at a high scale.
Thus, considerable attention has recently
been focussed on the renormalization group evolution~\cite{schrempp1}
of the various dimensionless Yukawa couplings in the SM .
Using the renormalization group evolution, one may attempt to relate the
Yukawa couplings to the gauge couplings via the Pendleton-Ross infra-red
stable fixed point (IRSFP) for the top-quark Yukawa coupling~\cite{pendross},
or via the quasi-fixed point behaviour~\cite{hill}.  The predictive power
of the SM and its supersymmetric extensions may, thus, be enhanced if
the renormalization group (RG) running of the parameters is dominated by
infra-red stable fixed  points.  
Typically, these fixed points are for ratios like Yukawa couplings
to the gauge coupling, or in the context of supersymmetric models, the 
supersymmetry breaking trilinear $A$-parameter to the gaugino mass, etc.
These ratios do not attain their fixed point values at the weak scale, the
range between the GUT (or Planck) scale and the weak scale being too small
for the ratios to closely approach the fixed point.  Nevertheless, the
couplings may be determined by quasi-fixed point behaviour~\cite{hill}
where the value of the Yukawa coupling at the weak scale is independent of
its value at the GUT scale, provided the Yukawa coupling at the GUT scale
is large.  For the fixed point or the quasi-fixed point scenarios to
be successful, it is necessary that these fixed points  are
stable~\cite{allanach,abel,jack}.  In the next section the above are
reviewed and in the last the results pertaining to 
a theory with R-parity and baryon and lepton number violation are
briefly summarized.

\section{Fixed points and stability}
In the SM, the top-quark Yukawa coupling $h_t$, satisfies the following one-loop
evolution equation:
\begin{eqnarray}
& \displaystyle 
8 \pi^2 {d h_t \over dt}=h_t \left( 9 g_3^2 +{3\over 4} (3 g_2^2+ g_1^2)+
{2\over 3} g_1^2 - {9\over 2} h_t^2\right), & 
\end{eqnarray}
where $t\equiv \ln(\mu_0^2/\mu^2)$, $\mu_0$ and $\mu$ are the initial
and final momentum scales for the evolution and $g_3,\, g_2,\ g_1$
are the gauge couplings corresponding to $SU(3),\, SU(2),\, U(1)$
respectively.  Ignoring all except the strong gauge coupling constant
$g_3$ allows us to write down the following equation for the ratio
of the Yukawa and gauge coupling:
\begin{eqnarray}
& \displaystyle
8 \pi^2 {d \ln(h_t/g_3) \over dt}= g_3^2 -{9\over 2} h_t^2. &
\end{eqnarray}
Setting the r.h.s. to zero yields the true fixed point for the top-quark
Yukawa coupling in the SM, $h_t^2=2 g_3^2/9$.  We note here that at
one-loop order, it is possible to discuss the solutions of the evolution
equations even if the electro-weak gauge couplings are retained.
In general we recall the following definitions:
\begin{eqnarray}
& \displaystyle {d g_i^2\over dt}=-{b_i\over 16 \pi^2} b_i g_i^4, \, \,
{d Y_t\over dt}= Y_t\left( \sum r_i \tilde{\alpha}_i - s Y_t \right),
\, i=1,2,3,&
\end{eqnarray} 
where $\tilde{\alpha}_i=g_i^2/(16 \pi^2), \, i=1,2,3 \, Y_t=h_t^2/(16 \pi^2)$.
It may be shown that since the range of evolution of the couplings is
finite the true fixed points are not reached in the infra-red regime,
and instead quasi-fixed solution emerges.  Furthermore, there is relative
insensitivity to the initial values of the Yukawa couplings, the focussing
property of the evolutions equations (this will be suitably
illustrated in the next section).  Retaining only the strong
gauge coupling yields the following result for the quasi-fixed point
at the scale $t$:
\begin{eqnarray}
& \displaystyle \left({Y_t\over \tilde{\alpha}_3}\right)^{QFP}=
{(Y_t/\tilde{\alpha}_3)^*\over 1-\left[\alpha_3(t)/\alpha_3(0)\right]^{B_3}}, &
\end{eqnarray}
where $B_3=r_3/b_3+1$ and defines the rate of attraction, and
$(Y_t/\tilde{\alpha}_3)^*=B_3 b_3/s$ is the true fixed point value.
In Table 1., we list the quantitites of interest for the SM and the MSSM.
\begin{table}
\begin{tabular}{||c|c|c||}
 &  SM & MSSM \\ \hline \hline
$b_3$ & $-7$ & $-3$ \\
$r_3$ & $8$ & ${16/ 3}$ \\
$s$ & ${9/2}$ & $6$ \\ 
$B_3$ & $-{1/7}$ & $-{7/ 9}$ \\
$(Y_t/\tilde{\alpha}_3)^*$ & ${2/9}$ & ${7/18}$ \\
\end{tabular}
{\centerline \caption{}}
\end{table}

It turns out that in the MSSM the quasi-fixed point analysis yields a
value for the top-quark mass that is in surprisingly good agreement with
the experimentally observed value and the rate of attraction to the true
fixed point value is better.  In practice, in the MSSM, the presence
of the parameter $\tan\beta$ which can be $\gg 1$ and can yield
phenomenologically desirable results~\cite{als}, naturally requires
us to consider a scenario where more than one Yukawa coupling can simultaneouly
reach fixed points.  This would quantitaively alter the predictions of
for the top-quark mass slightly, and raises the important question of
the stability of the fixed points of interest.  The analysis rests on
the stability of coupled first order differential equations and a thorough
analysis can now be performed.    In what follows, we shall only retain
the strong gauge coupling and consider the system of evolution equations
for the ratios $R_i=Y_i/\tilde{\alpha}_3$, with $i$ running over
all the Yukawa couplings which satisfy
the following evolution equations in a self-explanatory notation:
\begin{eqnarray}
& \displaystyle
{d R_i\over dt}=\tilde{\alpha}_3
R_i \left[(r_i + b_3) - \sum S_{ij} R_j \right],&
\end{eqnarray}
where $r_i = \sum_R 2 C_R$, and $C_R$ stands for the color quadratic Casimir
of the field.
Fixed points arise when if the $R_i^*=0$ or when
$R_i^*=\sum (S^{-1})_{ij} (r_j+b_3)$.  The stability of the solutions
may be tested by linearizing the system about the fixed points.
For the non-trivial fixed points we need to consider the
eigenvalues of the stability matrix whose elements are given by
$A_{ij}=R_i^* S_{ij}/b_3$.  The fixed point is infra-red stable
if the real part of each of the eigenvalues is negative.  For
the trivial fixed points the linear stability analysis requires
that each $\lambda_i=(\sum_j S_{ij} R_j^* -(r_i + b_3))/b_3$ also
be negative.  In practice, it turns out that all Yukawa couplings
that are associated with colored fields are required to attain non-trivial
fixed points, and those Yukawa couplings which are associated
with colored singlets attain trivial fixed points in order for the
fixed points to be infra-red stable, at least in models of interest.
The extension of the analysis above to the soft supersymmetry breaking
sector may also be done.  In particular, for the ratio of the
tri-linear couplings to the gaugino mass,
the analysis is particulary straightforward.  A close scrutiny of
the evolution equations for these reveals that in the event that some of
the Yukawa couplings attain non-trivial fixed points, the ratio
$A_i/m_{\tilde{g}}$ attains the value of 1, 
and values related to the values of fixed point Yukawa couplings
for all those that correspond to trivial fixed points.  Moreover, the
stability analysis need not be carried afresh and corresponds to
that of the Yukawa coupling sector.

\section{Models with R-parity and baryon and lepton number violation}
Here we will summarize briefly the results of some recent investigations
involving R-parity and Baryon and Lepton Number violation.  We recall that
R-parity is a discrete symmetry imposed on the MSSM to prevent rapid
proton decay through dimension 4 operators~\cite{weinberg}.  
One may add all terms consistent with gauge
symmetry and supersymmetry and introduce the following lepton
and baryon number violating terms to the
superpotential:
\begin{eqnarray}
& \displaystyle W_L= {1\over 2}\lambda_{abc} L^a_L L^b_L \overline{E}^c_R
+ \lambda'_{abc} L^a_L Q^b_L\overline{D}^c_R
+ \mu_i L_i H_2, & \label{Lviolating} \\
& \displaystyle W_B={1\over 2}\lambda''_{abc} \overline{D}^a_R
\overline{D}^b_R \overline{U}^c_R,  & \label{Bviolating} 
\end{eqnarray}
respectively, where $L,\, Q, \, \overline{E},\, \overline{D},\, \overline{U}$
denote the lepton and quark doublets, and anti-lepon singlet, d-type 
anti-quark singlet and u-type anti-quark singlet,  respectively.
Statistics and proton decay constraints
will restrict us to consider the following couplings
$\lambda_{233}, \lambda'_{333}, \lambda_{233}''$ only.    The fixed point
behaviour of these couplings has been considered in great detail 
in the recent past~\cite{ap1,ap2}.   It turns out that there is
only class of solutions that gives physically acceptable results
and is infra-red stable and corresponds to one where,
$R_\tau^*=0, \, R''^*=77/102, \, R_t^*=R_b^*=2/17$.  (The theory
may be enhanced to include one of the L violating coupling $R^*=0$
which remains indistinguishable from the first case.) 
It may also be seen in a straightforward manner that the
ratio of the tri-linear couplings to the gluino masses attain
their fixed point values: 
$\tilde{A}_{\lambda''}^*=\tilde{A}_b^*=\tilde{A}_t^*=1$, and
$\tilde{A}_\tau^*=-2/17$, where the latter is intimately related
to the values of the fixed points $R_b^*$.

\begin{figure}
\epsfxsize=8cm
\centerline{\epsfbox{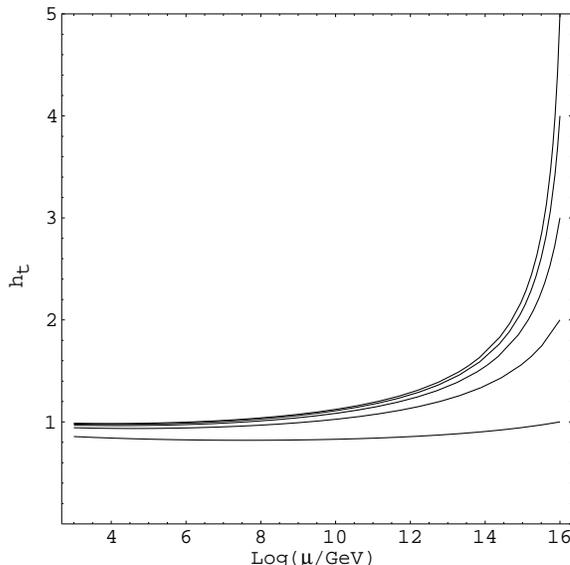}}
\caption{Renormalization group evolution of the
top-quark Yukawa coupling $h_t$ as a function of the logarithm
of the energy scale. We have taken the initial values of
$h_t$ at the scale $M_G \sim 10^{16}$ to be $5.0, \, 4.0, \, 3.0, \,
2.0$, and $1.0$.  The initial values of other Yukawa couplings 
are $h_b=0.91,\, h_\tau=0,\,$  and $\lambda''_{233}=1.08$.}
\label{fig:fig1}
\end{figure}

\begin{figure}
\epsfxsize=8cm
\centerline{\epsfbox{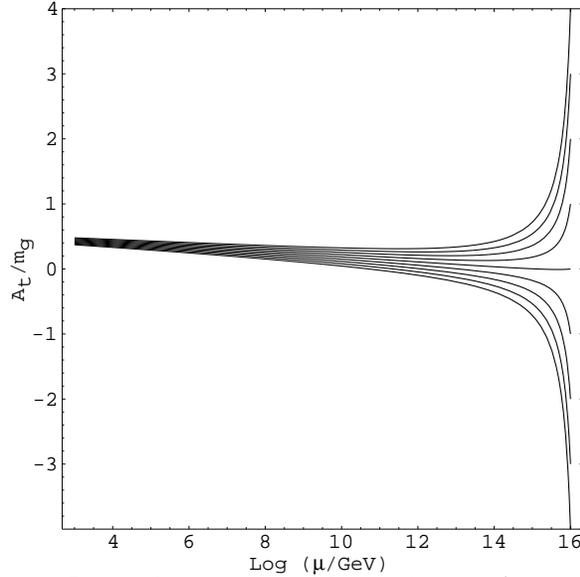}}
\caption{Renormalization group evolution of ratio 
$A_t/m_{\tilde{g}}$, as a function of the logarithm of the energy scale for 
several different initial values at $M_G$.
The initial values for other parameters at $M_G$ are
$h_t=5.0, \, h_b=0.91,\, h_\tau=0,\, \lambda''_{233}=1.08$,
and $A_b/m_g = 1.94, \, A_{\lambda''}/m_g = 2.57$.}
\label{fig:fig2}
\end{figure}

We finally illustrate several of the remarks made during the course of this
discussion in Figs.~\ref{fig:fig1} and ~\ref{fig:fig2}.  The focussing
property of the one-loop evolution equations for the Yukawa couplings
is brought out in Fig.~\ref{fig:fig1}, where the running Yukawa
coupling is plotted as a function of the logarithm of the
momentum scale.  We see that for a range of values
of the Yukawa coupling at the unification scale, the evolution equations
focus them to the neighbourhood of the quasi-fixed point.  This is also
true for the b-quark and baryon number violating coupling Yukawa couplings
$h_b,\, \lambda''_{233}$ which are not shown here (a complete discussion
can be found in ref.~\cite{ap2}).  We illustrate the behaviour of the
ratio $\tilde{A}_t$ plotted as a function of the logarithm of the
momentum scale.  The focussing property is also well illustrated in
this case.  In conclusion, we note that the
 presence of infra-red stable fixed points in supersymmetric models
provides a valuable guide to the possible range of parameters of the model
which are model independent.  In the MSSM the top-quark mass comes out in
the range predicted by these considerations.  R-parity and baryon and
lepton number violating models and the range of parameters therein,
can  be constrained to a narrow class that is favored by infra-red stable 
fixed points.

\end{document}